\documentclass[
aip,
apl,%
amsmath,amssymb,
preprint,
]{revtex4-1}

\usepackage{graphicx}
\usepackage{dcolumn}
\usepackage{bm}
\usepackage{amssymb,multirow,wrapfig}
\usepackage{color}
\usepackage{graphicx}
\usepackage{ulem}

\newcommand{\abs}[1]{\ensuremath{\left| #1 \right|}}

\newcommand{\ket}[1]{\ensuremath{\left| #1 \right>}}

\newcommand{\be}[0]{\begin{equation}}
  \newcommand{\ee}[0]{\end{equation}}
\newcommand{\bea}[0]{\begin{eqnarray}}
  \newcommand{\eea}[0]{\end{eqnarray}}

\begin{document}


\title{Compressive Object Tracking using Entangled Photons} 



\author{Omar S. Maga\~{n}a-Loaiza}
\email[]{omar.maganaloaiza@rochester.edu}
\affiliation{The Institute of Optics, University of Rochester, Rochester, NY 14627, USA}

\author{Gregory A. Howland}
\affiliation{Department of Physics and Astronomy, University of Rochester, Rochester, NY 14627, USA}

\author{Mehul Malik}
\affiliation{The Institute of Optics, University of Rochester, Rochester, NY 14627, USA}

\author{John C. Howell}
\affiliation{Department of Physics and Astronomy, University of Rochester, Rochester, NY 14627, USA}

\author{Robert W. Boyd}
\altaffiliation[Also at ]{Department of Physics, University of Ottawa, Ottawa, Ontario K1N 6N5, Canada.}
\affiliation{The Institute of Optics, University of Rochester, Rochester, NY 14627, USA}


\date{\today}

\begin{abstract}
  We present a compressive sensing protocol that tracks a moving
  object by removing static components from a scene. The
  implementation is carried out on a ghost imaging scheme to minimize
  both the number of photons and the number of measurements required
  to form a quantum image of the tracked object. This procedure tracks
  an object at low light levels with fewer than 3\% of the measurements
  required for a raster scan, permitting us to more effectively use the information content in each photon.
\end{abstract}

\pacs{}

\maketitle 

Compressive sensing (CS) has recently been of great utility in quantum
optical and low-light level applications, for instance, single-photon
level imaging, entanglement characterization and ghost
imaging\cite{Wang:2012vs,Yu:2012ud, Howland:2011cp, Howland:2013wf, Katz:2009fv}. CS provides a
resource-efficient alternative to single-photon arrayed detectors,
permitting us to reduce operational problems involved in systems
employing raster scanning\cite{Zerom:2011fu}.

CS applies optimization to recover a signal from incomplete or noisy
observations of the original signal through random
projections\cite{Candes:2006tb}. These ideas applied to the field of
imaging allow one to retrieve high resolution images from a small
number of measurements\cite{Candes:2008vs}. Recently, the quantum
optics community has employed CS for quantum state
tomography,\cite{Gross:2010cv,Liu:2012vd} to demonstrate nonclassical
correlations\cite{Howland:2013wf} and to form compressed ghost
images\cite{Zerom:2011fu}.

Ghost imaging is a technique which employs the correlations between
two light fields to reproduce an image. For example, entangled photons
exhibit strong correlations in many properties such as time-energy and position-momentum \cite{Jha:2008wu}. One photon of an
entangled pair illuminates an object and is collected by a bucket
detector, which does not provide spatial information. Its entangled
partner photon is then incident on a spatially resolving detector
gated by the first photon's bucket detector. Remarkably, an image of
the object appears on the spatially resolving detector, even though
its photon never directly interacted with the object
\cite{Pittman:1995jb}.

Compressive ghost imaging\cite{Katz:2009fv} allows one to replace the
spatially resolving detector with a bucket detector. This procedure
reduces both acquisition times for systems based on raster scanning
and the required number of measurements for retrieving
images\cite{Zerom:2011fu}. These improvements have motivated an
ongoing effort to implement technologies based on ghost imaging such
as image encryption \cite{Clemente:2010gu}, quantum sensors
\cite{Karmakar:2010fj}, object identification\cite{Malik:2010cr} and
most recently ghost imaging ladar\cite{Gong:2013wt}.

In spite of the advantages that technologies based on ghost
imaging offer, they can be hard to implement in practice. Most current
quantum optical technologies work at the single photon level, and are
unfortunately vulnerable to noise and are inefficient, requiring many
photons and many measurements \cite{OBrien:2009eu}. To reduce these
limitations, we apply an efficient form of compressive sensing. This
allows us to overcome the main problems which undermine the practical
application of many attractive correlated optical technologies. To
demonstrate these improvements, we implement a ghost object tracking
scheme that significantly outperforms traditional techniques. This opens the 
possibility of using correlated light in realistic applications for sparsity-based
remote-sensing.

We present a proof-of-principle experiment based on a quantum ghost
imaging scheme that allows us to identify changes in a scene using a small number of
photons and many fewer realizations than those
established by the Nyquist-Shannon criterion.  \cite{Shannon:1949fb}\
Object tracking and retrieval is performed significantly faster in
comparison to previous protocols
\cite{Malik:2011ws,Pittman:1995jb,Bennink:2002jr,Gatti:2004ec,Katz:2009fv,Zerom:2011fu}. This
scheme uses compressive sampling to exploit the sparsity of the
relative changes of a scene with a moving object. With this approach
we can identify the moving object and reveal its trajectory. Our
strategy involves removing static components of a scene and reduces
the environmental noise present during the measurement process. This
leads to the reduction of the number of measurements that we take
and the number of photons required to form an image, both important issues in proposals for object tracking and
identification \cite{Malik:2010cr,Malik:2011ws}. The reduction of
noise and removal of static components of a scene is carried out by
subtracting two observation vectors, corresponding to two realizations
of a scene. We call this technique ghost background subtraction. Our
results demonstrate that this technique is adequate for object
tracking at low light levels.

Consider the ghost imaging scheme depicted in Fig.  1.  A laser pumps
a nonlinear crystal oriented for type-I spontaneous parametric
down-conversion (SPDC). The approximated output state is given by
first order perturbation theory, which leads us to the following
two-photon entangled state:
\bea\ket{\Psi}&=&\int{d\vec{k}_{g}d\vec{k}_{o}f(\vec{k}_{g}+\vec{k}_{o}){\hat{a}^\dagger_{g}(\vec{k}_{g})}\hat{a}^\dagger_{o}(\vec{k}_{o})\ket{0}}.\eea

We refer to the down-converted photons as the ghost and object photons
denoted by the subindices $g$ and $o$, respectively. The two-photon
probability amplitude, which is responsible for the transverse
momentum correlations existing between the ghost and object photons,
is represented by the non-factorizable function
$f(\vec{k}_{g}+\vec{k}_{o})$, where $k$ is the transverse wavevector
of the ghost or object photon. The form of this function depends on
the phase-matching conditions, but it is often approximated by a
double gaussian function \cite{Monken:1998et}. This two-photon
entangled state is strongly anti-correlated in transverse momentum,
such that if the transverse momentum of the object photon is measured,
the transverse momentum of the ghost photon is found to have the same
magnitude and opposite direction. These momentum anti-correlations
allow us to perform ghost imaging.

\begin{figure}[t!]
  \centering\includegraphics[scale=.32]{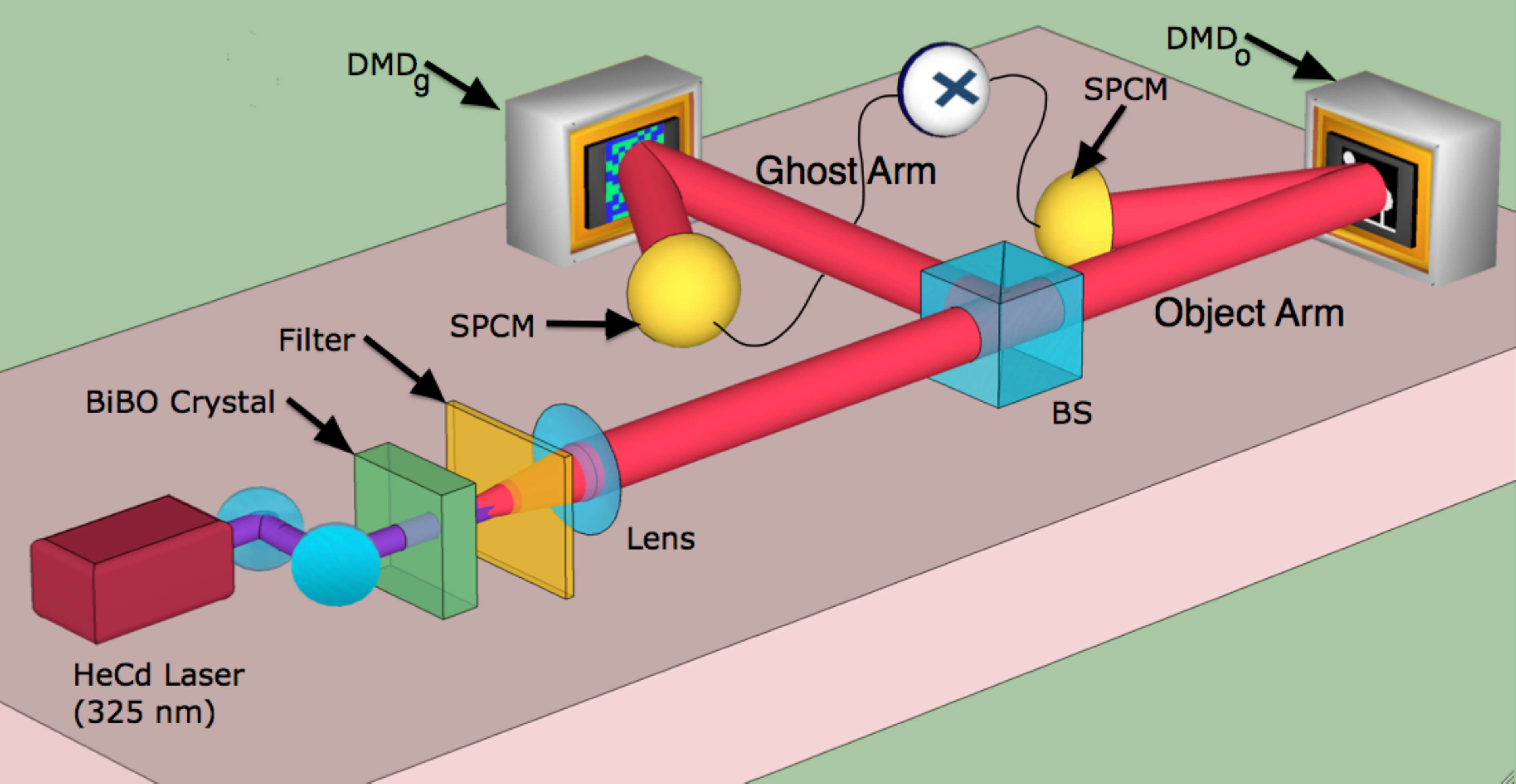}
  \caption{Entangled photons at 650 nm are generated in a Bismuth
    Barium Borate (BiBO) crystal through type-I degenerate spontaneous
    parametric downconversion (SPDC). The far field of the BiBO
    crystal is imaged onto two digital micromirror devices (DMDs) with
    a lens and a beam splitter (BS). One DMD is used to display the
    object we want to track, while the other is used to display random
    binary patterns. Single-photon counting modules (SPCMs) are used
    for joint detection of the ghost and object photons. }
  \label{SQIvQKD}
\end{figure}

In our experiment, we use digital micromirror devices (DMDs) to
impress spatial information onto the entangled photon pair. The DMDs
work by controlling the retro-reflection of each individual pixel on
the display. After each photon is reflected by a DMD, a single-photon
counting module (SPCM) counts the number of photons in it. The
correlations between the two down-converted photons allows one to
correlate the images displayed in the DMDs.

We jointly detect photons pairs reflected off a changing scene $O$ and
a series of random matrices $A_m$. The subindex $m$ indicates the
$m$-$th$ realization. The coincidence counts between the two detectors
are given by

\bea{J_m}\propto\int{d\vec{\rho}_{\tiny{_{\rm{DMD}}}}\abs{{A_m}\left(\frac{\vec{\rho}_{\tiny{_{\rm{DMD}}}}}{m_r}\right)}^2\abs{O\left(\frac{\vec{-\rho}_{\tiny{_{\rm{DMD}}}}}{m_o}\right)}^2},\label{eq:corr}\eea

where $A_m$ and $O$ are the reflectivity functions displayed on the
DMD$_g$ located in the ghost arm and on DMD$_o$ in the object arm,
respectively. Meanwhile $m_r$ and $m_o$ are their corresponding
magnification factors. These are determined by the ratio of the
distance between the nonlinear crystal to the lens and the distance
from the lens to DMD$_g$ or DMD$_o$.  In our experiment $m_r$ and
$m_o$, are equal. $\vec{\rho}_{\tiny{_{\rm{DMD}}}}$ represents the
transverse coordinates of one of the DMDs.

Eq.\( \, 2 \) critically shows that the joint-detection rate is
proportional to the spatial overlap between the images displayed on
DMD$_o$ and DMD$_g$.  This behavior can be interpreted as a nonlocal
projection, which demonstrates the suitability for implementing
compressive sensing techniques nonlocally with ghost
imaging\cite{Zerom:2011fu}.

Compressive sensing uses optimization to recover a sparse
$n$-dimensional signal from a series of $m$ incoherent projective
measurements, where the compression comes from the fact that $m<n$.
Image reconstruction via compressive sensing consists of a series
of linear projections \cite{Marcia:2011ve}. Each projection is the
product of the image $O$ consisting of $n$ pixels, with a pseudorandom
binary pattern $A_m$. Each pattern produces a single measurement,
which constitutes an element of the observation vector $J$. After a
series of $m$ measurements, a sparse approximation $\hat{O}$ of the
original image $O$ can be retrieved by solving the optimization
problem, known as total variation minimization\cite{CLi}, given by
Eq. 3.

\bea{min_{\hat{O}\in{C^{n}}}}\sum_i\abs{\abs{D_i\hat{O}}}_1+\frac{\mu}{2}\abs{\abs{A\hat{O}-J}}^2_2.\eea

$D_i\hat{O}$ is a discrete gradient of $\hat{O}$ at pixel $i$, $\mu$ is a weighting factor between the two terms, and $A$ is the total
sensing matrix containing all the pseudorandom matrices $A_m$. Each
matrix $A_m$ is represented into a 1D vector and constitutes a row of
the total sensing matrix $A$. The algorithm known as “Total Variation
Minimization by Augmented Lagrangian and Alternating Direction”
(TVAL3) allows us to solve the aforementioned problem. The solution of
the optimization problem allows us to recover the image $\hat{O}$,
which is the compressed version of the original image $O$, with a
resolution given by the dimensions of the matrix $A_m$. The original image O is characterized by a sparsity number $k$, which means that the image can be represented in a certain sparse basis where $k$ of its coefficients are nonzero. The number of
performed measurements $m$ is greater than the sparsity number $k$, but far fewer than the total number of
pixels $n$ contained in the original image. The constraints imposed
in the recovery algorithm minimize the noise introduced during the
measurement process.

We are able to compressively track and identify a moving object in a
scene by discarding static pixels. A scene with a moving object
possesses static elements that do not provide information about the
object's motion or trajectory. These redundancies can be discriminated
from the moving object as follows. Let us consider the projection of
two different frames onto the same pseudorandom pattern. Each
projective measurement picks up little information about the
components of a frame. If the two projective measurements produce the
same correlation value, it would imply that the two frames are
identical and we are retrieving meaningless information which can be
ignored. The opposite case would reveal information about the changes
in a scene.

This protocol is formalized as follows. Two different
correlation vectors, $J^j$ and $J^{j-1}$, corresponding to two
consecutive frames are subtracted, giving $\Delta J$. This introduces
the following important modification to Eq. 3.
\bea{min_{\hat{O}\in{C^{n}}}}\sum_i\abs{\abs{D_i{\Delta}\hat{O}}}_1+\frac{\mu}{2}\abs{\abs{A{\Delta}\hat{O}-{\Delta}J}}^2_2.\eea

\noindent The subtracted vector ${\Delta}J$ is sparser than both $J^j$
or $J^{j-1}$, thus requiring fewer measurements for its
reconstruction. This corresponds to fewer realizations of $A_m$, and
hence smaller sensing matrix $A$.  Furthermore, subtracting the
background in this manner mitigates the environmental noise present
during the tracking process. The retrieved image ${\Delta}\hat{O}$
will provide information about the relative changes in the scene.

Our experimental setup is sketched in Fig. 1. A 325 nm,
continuous-wave HeCd laser pumps a type-I phase matched BiBO crystal
to produce degenerate entangled photon pairs at 650 nm. Two
interference filters are placed after the nonlinear crystal. The first
is a low pass filter that removes the pump and the second is a 650/12
nm narrowband filter that transmits the down-converted photons. A beam
splitter probabilistically separates the two photons into ghost and
object modes. An 88 mm focal length lens puts the far field of the
crystal at the location of DMD. Two free space detectors receive the
light reflected from the DMDs by means of two collection lenses with a
25 mm focal length. One DMD is used to display a scene with a moving
object while the other is used to impress a series of random binary
patterns. Coincidence counts are obtained within a 3 ns time window.

\begin{figure}[b!]
  \centering\includegraphics[scale=.38]{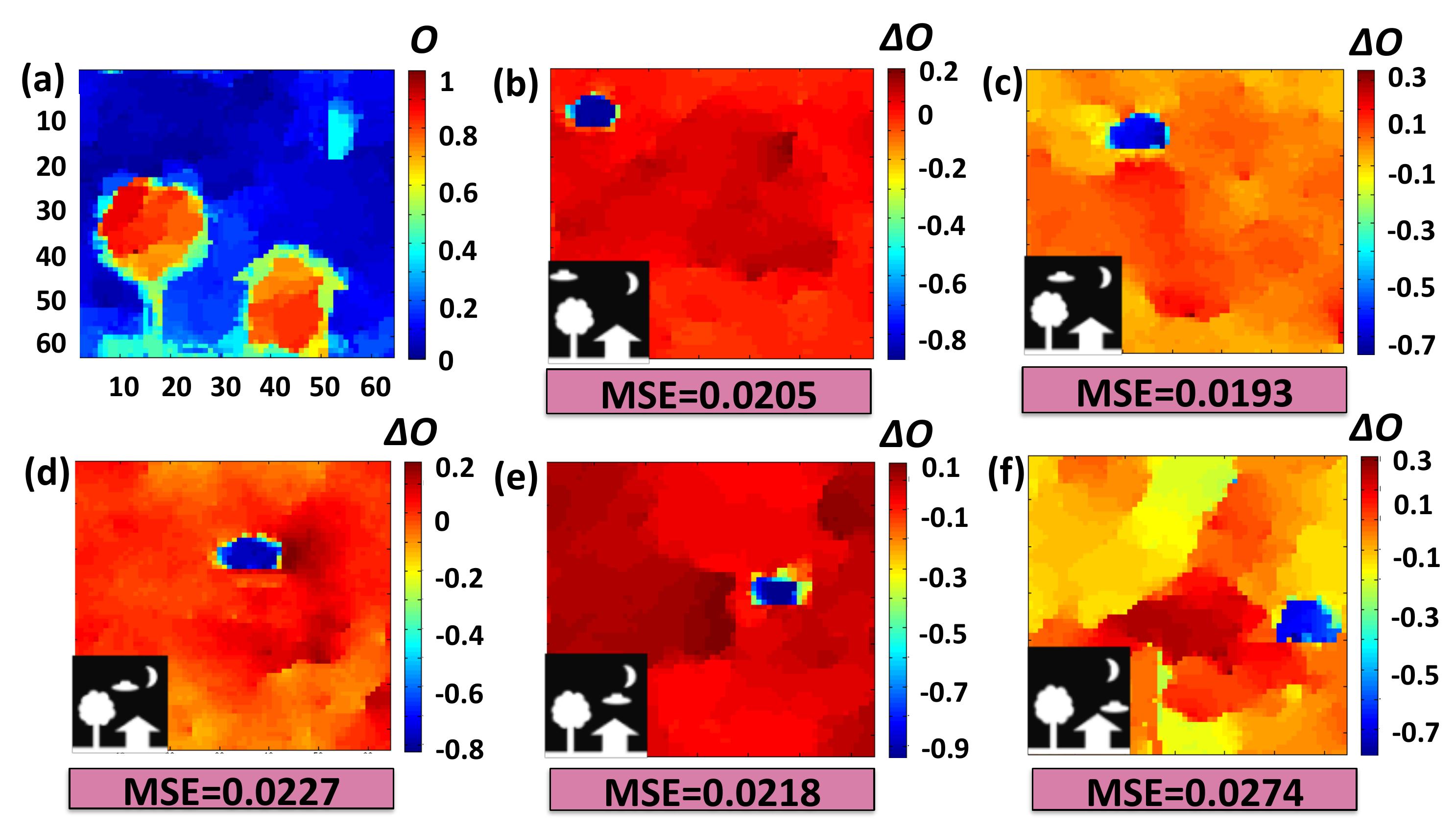}
  \caption{Compressed ghost image of (a) the background of the scene and
    (b-f) the tracked object in different positions. These
    reconstructions were obtained by defining different ${\Delta}J$
    vectors with 400 elements, corresponding to the number of
    measurements. The insets show the original frames of the scene displayed
    on the DMD.  }\label{ghostimage}
\end{figure}

We apply this method to a scene with a flying
object. The static components of the scene are a house,
the moon and a tree. The object moves a certain distance in each iteration of the scene (insets of Fig. \ref{ghostimage}). We first reconstruct a compressed
ghost image of the static frame of the scene, which represents the background. In order to do this, we
put 2000 different random patterns on
DMD$_g$, with DMD$_o$ displaying the background scene. These realizations
represent 49\% of a raster scan.  For each random pattern, we count coincidence
detections for 8 s. Typical single count rates were
$13.8$ x $10^3$ counts/s for the ghost and object arms with the coincidence
counts approximately 2\% of the single counts. Fig. 2(a) shows the retrieved background scene $\hat{O}$. After this, subsequent frames of the scene with the object in different positions are displayed on DMD$_o$. After applying the optimization algorithm, the moving object was clearly identified as shown in Figs. 2(b)-(f). The reconstructions were done using 400 patterns, which represents 9.7\%
the measurements of a raster scan. The negative values in the
retrieved images are due to background subtraction and
fluctuations in the measurements process.

A straightforward examination of the limits of our protocol is carried out by reducing the number of measurements used to
track an object. The images shown in Fig. 3 were reconstructed with
only 200 and 100 measurements, corresponding to 4.88\% and 2.44\% of the
measurements of a raster scan. The metric employed to characterize the fidelity of
these reconstructions is the mean-squared error\cite{Katz:2009fv}
defined as $MSE=(1/n)\lVert{O-\hat{O}}\rVert ^2$. The $MSE$ is seen to increase as the number of measurements is decreased. Although, it is
still possible to detect the object trajectory with just 100 measurements.

The photon efficiency is studied by estimating the dependence of the
$MSE$ on the number of photons per measurement, for a fixed number
of measurements. A simulation of the protocol was carried out by using the data employed in the experiment. In order to achieve realistic experimental conditions, dark and shot noise were introduced by means of poissonian distributions. The amount of dark noise was modeled based on the frequency distribution of counts obtained when both of the DMDs were turned off.
We have considered reconstructions employing 100 and 400 measurements. Fig. 4 shows the dependence of image quality on the number of detected photons per measurement. The minimum number of photons per measurement
needed to distinguish the silhouette of the object by eye are 500 photons/measurement and 200
photons/measurement for 100 and 400 measurements respectively. The estimated thresholds correspond to a $MSE$ oscillating around 0.04. For the situation where an object was tracked with 100 measurements and 500 photons/measurement, we estimate that we can impress approximately 0.082 bits/photon. This is considering that for a binary image the number of pixels corresponds to the number of bits \cite{Nakadate:1980vb}.

\begin{figure*}[t!]
  \centering\includegraphics[scale=.56]{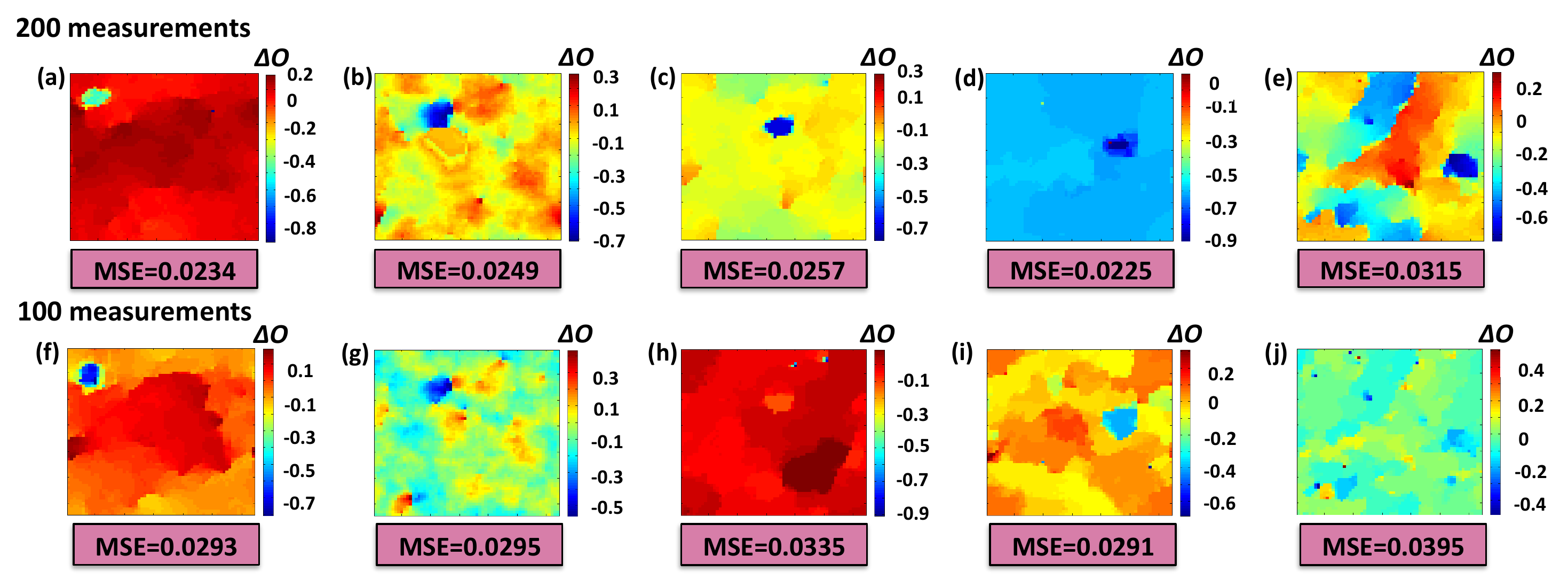}
  \caption{Reconstructed ghost image of (a-e) tracked object with 200
    measurements. (f-j) same object with 100
    measurements. }\label{pol_data}
\end{figure*}

The maximum object velocity that we can track is limited by the number of photons that we are able to detect. In our setup, each scene reconstruction took 13.3 minutes (for the case of 100 measurements) due to the low photon flux. If we were to use a high brightness source of entangled photons, we could shorten the acquisition time needed to retrieve a compressed ghost image with an $MSE$ below the threshold shown in Fig. 4. As such, there is no hard theoretical limit on the maximum object velocity that can be tracked using this method.

\begin{figure}[h!]
  \centering\includegraphics[scale=.43]{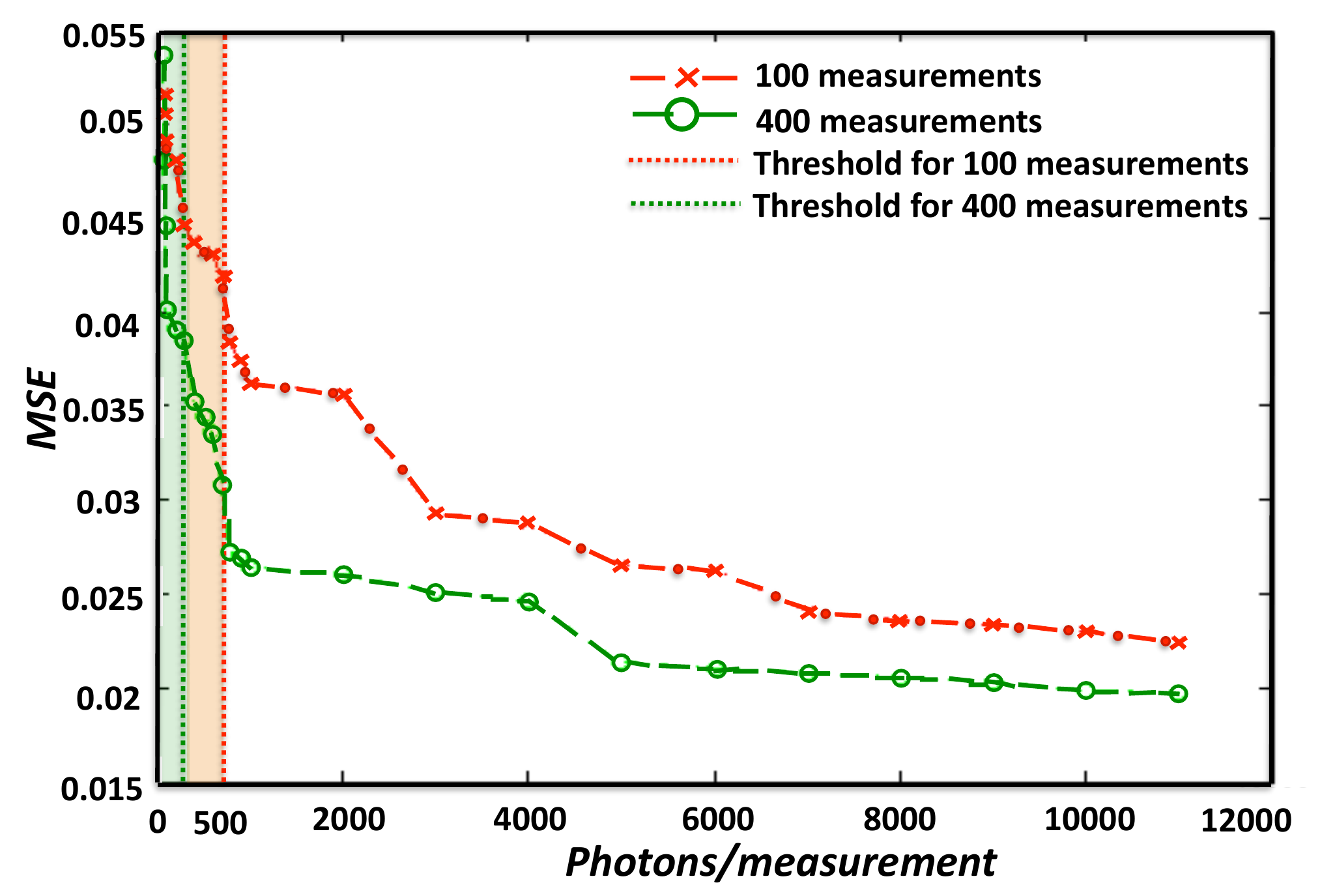}
  \caption{(Color online) Calculated mean-squared error of the
    compressed tracked object at the position shown in Fig. 2(b). Green
    (Red) line indicates the MSE using 400 (100) measurements. The
    thresholds indicate that a low quality image is retrieved and is not
    possible to track the object.  }\label{SQIvQKD}
\end{figure}

In conclusion, we have proposed and demonstrated a proof-of-principle object-tracking protocol in a ghost imaging scheme. This protocol uses compressive sensing to exploit the sparsity existing between two realizations of a scene with a moving object. It also reduces the
environmental noise introduced during the measurement process. Further, it allows us to perform image retrieval significantly faster by employing single
pixel detectors. Our method is photon-measurement efficient, allowing us to track an object with only
2.44 \% of the number of measurements established by the Nyquist criterion,
even at low light levels. This economic procedure shows potential for
real-life applications.

\begin{acknowledgments}
 The authors would like to thank M. Mirhosseini and A.C. Liapis, P. Ybarra-Reyes and J.J. Sanchez-Mondragon for helpful discussions. This work was supported by the DARPA AFOSR GRANT FA9550-13-1-0019 and the CONACYT.
\end{acknowledgments}

\end{document}